# IAPS | Institute for AI Policy and Strategy

April 27th, 2026

# Risk Reporting for Developers' Internal AI Model Use

**AUTHORS**

Oscar Delaney – Research Associate[†]

Sambhav Maheshwari – Research Associate

Joe O'Brien – Researcher

Theo Bearman – Researcher

Oliver Guest – Research Manager

[†] Work done while at IAPS

# Abstract

Frontier AI companies first deploy their most advanced models internally, for weeks or months of safety testing, evaluation, and iteration, before a possible public release. For example, Anthropic recently developed a new class of model with advanced cyberoffense-relevant capabilities, Mythos Preview, which was available internally for at least six weeks before it was publicly announced. This internal use creates risks that external deployment frameworks may fail to address.

Legal frameworks, notably California's Transparency in Frontier Artificial Intelligence Act (SB 53), New York's Responsible AI Safety And Education (RAISE) Act, and the EU's General-Purpose AI Code of Practice, all discuss risks from internal AI use. They require frontier developers to make and implement plans for how to manage risks from internal use, and to produce internal use risk reports describing their safeguards and any residual risks. This guide provides a harmonized standard for companies to produce internal use risk reports suitable for all three regulatory frameworks. It is addressed primarily to evaluation and safety teams at frontier AI developers, and secondarily to regulators and auditors seeking to understand what good reporting looks like.

Given the pace of AI R&D automation and the limited external visibility into how companies use their most capable models internally, regular and detailed risk reporting may be one of the few mechanisms available to ensure that the risks from internal AI use are identified and managed before they materialize. Whenever a substantially more capable or riskier model is deployed internally, the developer should create a risk report and argue why the model is safe to deploy. We structure the reporting framework around two threat vectors—autonomous AI misbehavior and insider threats—and three risk factors for each: means, motive, and opportunity.

# Executive Summary

Frontier AI companies first deploy their most advanced models internally, for weeks or months of safety testing, evaluation, and iteration, before a possible public release. For example, Anthropic recently developed a new class of model with advanced cyberoffense-relevant capabilities, Mythos Preview, which was available internally for at least six weeks before it was publicly announced.[1] This internal use[2] creates risks that external deployment frameworks may fail to address.

Legal frameworks, notably California's Transparency in Frontier Artificial Intelligence Act (SB 53), New York's Responsible AI Safety And Education (RAISE) Act, and the EU's General-Purpose AI Code of Practice, all discuss risks from internal AI use. They require frontier developers to make and implement plans for how to manage risks from internal use, and to produce internal use risk reports describing their safeguards and any residual risks.[3] This guide provides a harmonized standard for companies to produce internal use risk reports suitable for all three regulatory frameworks. It is addressed primarily to evaluation and safety teams at frontier AI developers, and secondarily to regulators and auditors seeking to understand what good reporting looks like.

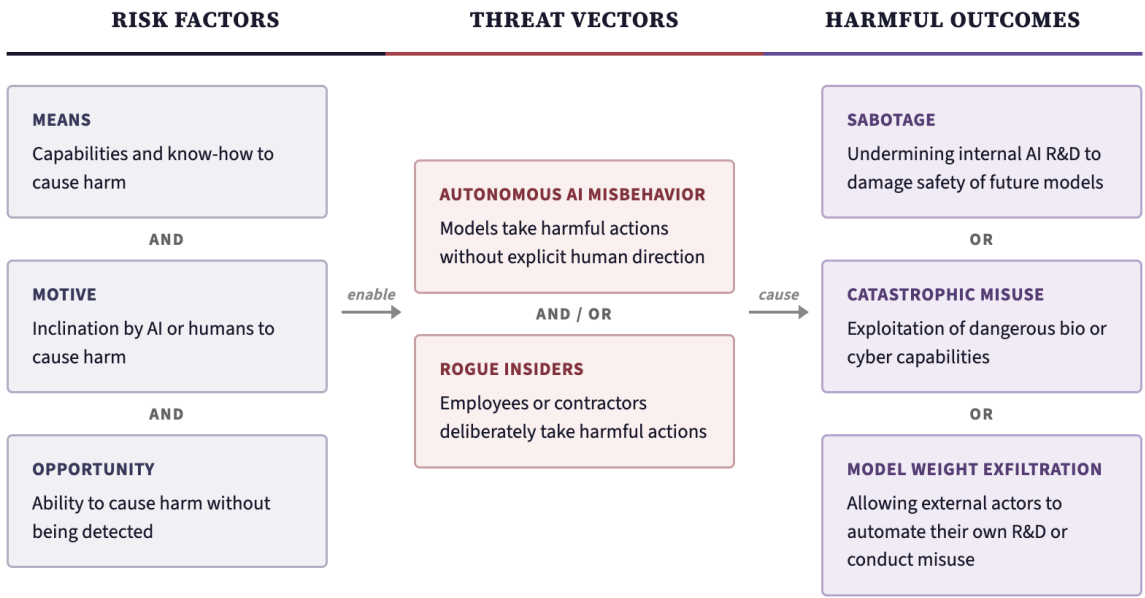

Figure 1: Risk taxonomy for internal AI model use. The conjunction of all three risk factors (left) and one or both threat vectors (center) can lead to any of the harmful outcomes (right).

---

[1] Anthropic, "Project Glasswing: Securing critical software for the AI era," announcing Mythos Preview, was published on April 7, 2026; Anthropic, "System Card: Claude Mythos Preview," 13, confirms that an early version of Mythos Preview was available internally at Anthropic from February 24, 2026.
[2] In this document, "internal use" refers to both unreleased models and internal deployments of publicly released models that feature enhanced capabilities, expanded access, or reduced safety filters.
[3] Note that the Code of Practice only applies to signatories (although other frontier developers still need to demonstrate compliance with the AI Act). SB 53 and RAISE require companies to describe their risk mitigation practices and comply with the plan they outline, but there is no minimum standard for this plan.

Figure 1 summarizes the risk landscape for internal AI models.[4] Given the pace of AI R&D automation and the limited external visibility into how companies use their most capable models internally, regular and detailed risk reporting may be one of the few mechanisms available to ensure that the risks from internal AI use are identified and managed before they materialize. Whenever a substantially more capable or riskier model is deployed internally, the developer should create a risk report and argue why the model is safe to deploy. Table 1 summarizes some key indicators the report should include for each combination of threat vector and risk factor.

## Key Internal Use Risk Indicators

| Threat Vector | Means | Motive | Opportunity |
| --- | --- | --- | --- |
| **Autonomous AI misbehavior (Section 4)** | Advanced capability benchmarks; AI R&D and software engineering evals; real-world R&D contribution metrics; covert reasoning assessments | Behavioral "honeypot" evaluations; reward hacking incidents; interpretability results; observed misbehavior logs | Internal use cases and permissions; monitoring and oversight systems; security testing; control evaluations |
| **Insider threats (Section 5)** | Uplift assessments vs. public models; dangerous capability benchmarks (bio, cyber, chemical); jailbreak resistance | Security vetting processes; security violation records; insider risk management statistics | Access controls and counts by tier; monitoring and logging; prevention mechanisms; social engineering defenses |

Table 1: Key internal use risk indicators to report. The reporting focus can generally be limited to the most capable or highest risk internal model. Insider threat indicators (bottom row) are primarily intended for confidential submission to regulators rather than public summary.

---

[4] Non-deliberate capability failures—such as models producing incorrect code because a task exceeds their abilities—are important but outside the scope of this report, which focuses on intentional or goal-directed risks.

# Table of Contents

# 1. Why Internal Models Pose Distinctive Risks

Frontier AI companies' internal models pose risks distinct from, and sometimes greater than, risks from externally deployed systems.[5]

**Internal models have privileged access to sensitive systems.** Internal models—and the employees who use them—often have access to sensitive systems such as training infrastructure, safety evaluation pipelines, model weights, security controls, and proprietary codebases. As these models become integrated into critical workflows, such as code review, security monitoring, and research, the attack surface they create expands correspondingly.

**Internal models lack external oversight.** Models used exclusively within a company are developed and tested with limited external scrutiny. The International AI Safety Report (2025) noted that "very little is publicly known about internal deployments."[6] Some models may never be released publicly at all, meaning they never receive the external red teaming, auditing, and public scrutiny that accompany a product launch.[7]

**Internal models are more capable than public models.** The most capable systems at any given company are typically deployed internally for weeks or months before public release.[8] OpenAI's GPT-4, for instance, was used internally for roughly six months before its public launch.[9] More recently, however, this internal-only period has shortened; GPT-5, for example, underwent just three weeks of safety testing by contractors.[10] Anthropic's Claude Mythos Preview—released to a small set of cybersecurity partners in April 2026—was deployed internally from late February 2026. However, Anthropic does not plan to make the model generally commercially available, on the grounds that its autonomous zero-day vulnerability discovery and exploitation capabilities would—if broadly available—accelerate cyberoffensive activity against major operating systems and browsers.[11] Greater capabilities entail greater potential for both accidental and intentional harm.

---

[5] Acharya and Delaney, "Managing Risks from Internal AI Systems."; Chan, "AI Models Can Be Dangerous Before Public Deployment."; Stix et al.,"AI Behind Closed Doors: a Primer on The Governance of Internal Deployment."; Kwon and Casper, "Internal Deployment Gaps in AI Regulation."
[6] Bengio et al., "International AI Safety Report 2025," 35.
[7] Safe Superintelligence Inc., "Safe Superintelligence Inc."
[8] Delaney and Acharya, "The Hidden AI Frontier."
[9] E.g., GPT-4 had a six month period of internal use and safety testing. See OpenAI, "GPT-4 Technical Report," 59. More recent models likely have had shorter internal use periods, though the exact length is often unknown.
[10] Note that there may have been a longer testing period internally—the report only says that METR was given access to the model for three weeks before public deployment. See OpenAI, "OpenAI GPT-5 System Card," 41.
[11] Anthropic, "Project Glasswing: Securing critical software for the AI era," announcing Mythos Preview, was published on April 7, 2026; Anthropic, "System Card: Claude Mythos Preview," 13, confirms that an early version of Mythos Preview was available internally at Anthropic from February 24, 2026.

Furthermore, insiders at frontier companies may use or develop helpful-only model variants—model versions without the safety filters typically applied to public-facing models. This practice substantially increases the potential for misuse.[12]

These features will be exacerbated as AI models become more deeply integrated into automated AI R&D:

- Companies may become more reluctant to publicly deploy models that materially speed up AI R&D, lest they help their competitors.[13] This reluctance could widen the gap between internal and public AI capabilities, making internal models more dangerous, less visible, and more attractive targets for exfiltration by adversaries.[14]
- The superhuman pace of automated internal R&D could strain human oversight capacity.
- AI models intended to automate AI research will need extensive access to company codebases, training infrastructure, and compute to run experiments. This privileged access increases the potential damage from model misbehavior.[15]

Consequently, the internal use of AI models for automated R&D is a critical focus for risk reporting. Detailed reporting on the scope of AI R&D automation can provide governments and the public with a leading indicator of accelerating AI capabilities.

## Threat Vectors and Harmful Outcomes

There are two main threat vectors from internal models, both recognized in the Frontier AI Safety Commitments agreed at the AI Seoul Summit[16] and reflected in the safety frameworks of leading AI developers, including Anthropic,[17] Google DeepMind,[18] and OpenAI:[19]

---

[12] For a recent example of helpful-only variants used in practice, see Anthropic, "System Card: Claude Mythos Preview," §1.1.4 (noting that "helpful only" snapshots without safeguards exist alongside the standard model during training) and §2.2.5.2 (deploying a helpful-only Mythos Preview snapshot in a virology protocol uplift trial against Opus 4.6-assisted and unassisted controls). The system card also reports running safeguard evasion and influence-operation evaluations against helpful-only variants to isolate raw capabilities (§§4.4 and 8.3).
[13] Anthropic, "Detecting and preventing distillation attacks."; Robinson, "Anthropic Revokes OpenAI's Access to Claude."
[14] Acharya and Delaney, "Managing Risks from Internal AI Systems."
[15] Acharya and Delaney, "Managing Risks from Internal AI Systems."
[16] The Frontier AI Safety Commitments (signed by 16 major AI companies in Seoul) explicitly require participants to "assess the risks posed by their frontier models or systems across the AI lifecycle, including before deploying that model or system, and, as appropriate, before and during training." See Department for Science, Innovation, and Technology, "Frontier AI Safety Commitments, AI Seoul Summit 2024."
[17] Anthropic lists AI systems engaging in autonomous internal sabotage as a potential threat. See Anthropic, "Responsible Scaling Policy," (v3.0, Feb. 24, 2026), § 1 (capability threshold for "High-stakes sabotage opportunities"); Anthropic, "Risk Report: February 2026," § 2.1.
[18] Google describes illustrative mitigations for "high-stakes internal deployments where there is significant risk of the model undermining human control." See Google DeepMind, "Frontier Safety Framework Version 3.0."
[19] OpenAI notes that safeguards "need to be robust to both malicious actors (either internal or external) and model misalignment risks." See OpenAI, "Preparedness Framework," 13.

- **Autonomous AI misbehavior.** Models may take harmful actions on their own initiative. This may include subtle interference with evaluations, such as sandbagging (deliberately underperforming to hide capabilities) or alignment faking (behaving well during evaluations while pursuing different goals during deployment). This could also include covert actions, such as models attempting to exfiltrate their own weights or sabotage the training of next-generation models.[20]
- **Insider threats.** Employees or contractors with privileged access may misuse, exfiltrate, or maliciously modify internal models, potentially in collaboration with external actors such as foreign governments or criminal organizations.[21] Internal AI models amplify traditional insider threats in two ways: they provide powerful tools that can dramatically increase the damage a malicious insider can inflict, and the models themselves are high-value targets for theft or sabotage.

The principal harmful outcomes that both threat vectors can produce fall into three broad categories:

- **Sabotage of internal AI R&D.** A rogue human or AI model could undermine safety and alignment research, or manipulate the training of subsequent generations of AI models (for example, by poisoning synthetic pretraining data, reinforcement learning environments, evaluations, or alignment data they help produce, or by corrupting the training code or AI-based evaluators used in the pipeline). These actions could compound over time and may be difficult to detect.[22]
- **Direct catastrophic misuse.** Advanced internal model capabilities could be used by people within AI companies for bioweapons development, large-scale cyberattacks, or other actions causing widespread harm.[23]
- **Model weight exfiltration.** A rogue human or AI could make and distribute an unauthorized copy of the model weights. Unlike sabotage and misuse, which involve harm from models being used *internally*, exfiltration removes the model from the developer's control entirely, enabling adversaries to fine-tune away safety measures, automate their own AI R&D, or conduct large-scale misuse without any of the developer's safeguards.[24]

---

[20] Anthropic, "Risk Report: February 2026."; Anthropic, "Sabotage Risk Report: Claude Opus 4.6."
[21] Outsider threats—such as external hackers or state-sponsored actors seeking direct unauthorized access to internal models—are relevant to the overall security posture of frontier AI companies. But stealing an internally-deployed model is likely no easier (and possibly harder) than stealing the weights of a publicly deployed model. So we do not focus on outsider threats, except via collusion with insider threats, which is addressed in Section 5.
[22] Greenblatt et al., "AI Control: Improving Safety Despite Intentional Subversion."
[23] Shevlane et al., "Model evaluation for extreme risks."; Martin, "Overview of Transformative AI Misuse Risks: What Could Go Wrong Beyond Misalignment."
[24] Nevo et al., "Securing AI Model Weights."; Qi et al., "Fine-tuning Aligned Language Models Compromises Safety, Even When Users Do Not Intend To!"

# 2. Legal Basis for Internal Use Risk Reports

Three major AI regulatory frameworks in California, New York, and the EU discuss risks from internal use. AI developers are required to produce **internal use risk reports** describing their safeguards and residual risks not sufficiently addressed by safeguards.

**California's SB 53** and **New York's RAISE Act**, which have substantially identical text, require developers of frontier AI models to write and publish (with necessary redactions) a frontier AI framework, including provisions addressing risks from internal use. Specifically, each frontier developer must describe its approach to:

> *Assessing and managing catastrophic risk resulting from the internal use of its frontier models, including risks resulting from a frontier model circumventing oversight mechanisms.*[25]

Additionally, frontier developers should review the "adequacy of mitigations as part of the decision to deploy a frontier model or use it extensively internally."[26] They must also report on their internal use risks "every three months or pursuant to another reasonable schedule."[27]

The **EU framework** has less clear-cut applicability, but arguably addresses internal use through several provisions. The EU AI Act requires AI providers to assess and mitigate risks from the development and use of general-purpose AI models with systemic risk,[28] though it is not specified explicitly whether internal use is also in scope.[29] The Code of Practice for General-Purpose AI Models, written with extensive input from AI companies, clarifies how AI providers can achieve compliance with the AI Act. It defines "use" as including "use of the model by the Signatory or other actors" (Glossary), includes "capabilities to automate AI research and development" among the sources of systemic risk that signatories must assess (Appendix 1.3.1), and requires signatories to "protect against insider threats, including in the form of (self-)exfiltration or sabotage carried out by models" (Appendix 4.4).[30] Moreover, signatories of the Code of Practice must provide the European AI Office with "a description of how the model has been used and is expected to be used, including its use in the development, oversight, and/or evaluation of [other] models" (Measure 7.1) and "a description of all security mitigations implemented" (Measure 7.3).[31]

---

[25] Section 22757.12(a)(10) SB 53 and Paragraph § 1421.1(j) RAISE Act. See California State Legislature, "SB-53 Transparency in Frontier Artificial Intelligence Act."
[26] Section 22757.12(a)(4) SB 53 and Paragraph 1421.1(d) RAISE Act.
[27] Section 22757.12(d) SB 53 and Paragraph 1422.2 RAISE Act.
[28] EU AI Act Article 55.
[29] For a detailed legal analysis see Pistillo, "Internal Deployment in the EU AI Act."
[30] Samwald et al., "Code of Practice for General-Purpose AI Models."

**AI developers are left with significant discretion about exactly what to include in internal use risk reports,** especially in the minimal SB 53 and RAISE laws. **This guide fills that gap by offering a harmonized reporting standard compatible with all three legal frameworks.**[32] It aims to set out gold-standard reporting practices: recommendations that go beyond the legal minimum in several respects, drawing on and extending existing industry practice. Many of the recommendations in Sections 4 and 5 reflect assessments that leading developers such as Anthropic have published in their risk reports;[33] formalizing them into a reporting template promotes consistency and enables cross-company comparison. Where this guide recommends reporting that goes beyond current industry practice—particularly around insider threats—it does so because these are safety gaps, and justifies these recommendations on safety grounds in the relevant sections.

This guide focuses on periodic risk reports rather than incident reporting, but developers should also ensure their incident reporting processes follow standards and best practices.[34] Finally, developers and regulators should note that the two threat vectors addressed in this guide have different disclosure sensitivities. Information about insider threat mitigations (Section 5) may itself provide a roadmap for malicious insiders if made public; this guide therefore recommends that Section 5 reporting be submitted to regulators in a confidential annex, separately from the AI misbehavior reporting in Section 4, which is more suitable for public summary.

---

[31] Samwald et al., "Code of Practice for General-Purpose AI Models."
[32] Note that this is not legal guidance, and AI developers should seek legal counsel on compliance with any relevant legislation!
[33] Anthropic, "Risk Report: February 2026."
[34] SB 53 (Section 22757.13) requires reporting critical safety incidents within 15 days, or within 24 hours if there is an imminent risk of death or serious injury. Moreover, the Code of Practice (Measure 9.3) requires serious incident reporting within 2 to 15 days depending on severity.

# 3. Risk Factors and Reporting Scope

Having established *why* internal use poses distinctive risks (Section 1) and the *legal basis* for reporting on them (Section 2), this section turns to *how* companies should structure their reports. It introduces the risk taxonomy used in Sections 4 and 5, and then discusses the scope and standards for internal use risk reports.

## 3.1 The Risk Framework

AI developers should aim to build a "safety case," i.e., a structured argument, supported by evidence, that the AI system is acceptably safe.[35] A safety case need not be a formal proof of safety; rather, it is a comprehensive assessment of the available evidence, organized to make the reasoning transparent and contestable.[36] This approach aligns with the EU Code of Practice's requirement for a "systemic risk acceptance determination" with "detailed justification for why the systemic risks stemming from the model are acceptable" (Measure 7.2). For risks from internal models, this requires separate arguments for the autonomous AI misbehavior and the rogue insider threat pathways. To make a strong safety case, developers should discuss each of three risk factors across both autonomous AI and human insider threat models (see Table 1):[37]

- **Means:** Does the model or insider have the technical capabilities to cause the relevant harm, including the ability to carry out the harmful action covertly, evading monitoring and oversight systems?[38]
- **Motive:** Is the model or insider inclined to cause harm?
- **Opportunity:** Could the model or insider get the chance to cause harm undetected, given the access controls, permissions, monitoring, and other safeguards in place?

Safety arguments that rest on the absence of **means** are the most likely to erode over time, as models become more capable. Arguments relying on **opportunity**—that access controls and monitoring would prevent or detect harmful actions—are strongest when regularly validated through adversarial testing, since controls can fail or be circumvented in unforeseen ways. In the long run, as models become more capable and more deeply integrated into critical systems, **motive**-based safety arguments will likely become the most important. Sufficiently capable and broadly deployed models may have both the means and the opportunity to cause harm should they choose to, and relying solely on controls and safeguards to rein in a powerful misaligned model would be a fragile defense.[39] To ensure defense in depth, AI developers should provide evidence and arguments across means, motive, and opportunity, mitigating the risk if new evidence undermines any single component.

---

[35] Clymer et al.,"Safety Cases: How to Justify the Safety of Advanced AI Systems."
[36] Note that "safety case," as used in some industries, may refer to a higher standard of proof.
[37] This three-factor risk model is drawn from METR, forthcoming.
[38] For a model, this may include stealth-forward capabilities like steganography (Zolkowski et al., "Early Signs of Steganographic Capabilities in Frontier LLMs.") or latent-space reasoning (Hao et al., "Training Large Language Models to Reason in a Continuous Latent Space.").

Importantly, a safety case for internal deployment does not require demonstrating that a model poses zero risk. Rather, it should argue that the internal deployment does not pose meaningfully greater *marginal* risk relative to publicly available AI systems and other tools.[40] This framing acknowledges that some baseline risk already exists from publicly deployed models, and focuses the safety case on the additional risk created by the internal deployment's enhanced capabilities, expanded access, or reduced safety filters—exactly the features that distinguish internal use from public deployment. Many of the evidence categories in Sections 4 and 5 are therefore most informative when reported as comparisons against publicly available models.

## 3.2 Reporting Coverage and Cadence

AI developers should focus on reporting risks from their highest-risk internal systems, with the comprehensiveness of reporting scaling with the magnitude of model capability advances since the last risk report. Quarterly reports serve as a backstop in case many small changes collectively constitute a large update to the risk profile. **Figure 2** presents the decision logic by which AI developers should determine what reports to make to the relevant regulator about their internal AI use.

---

[39] Greenblatt et al., "AI Control: Improving Safety Despite Intentional Subversion," 17.
[40] For further discussion of marginal risk as the appropriate standard for safety assessments, see Anthropic, "Responsible Scaling Policy," which frames risk thresholds in terms of the incremental risk a model poses over and above what is achievable with publicly available tools.

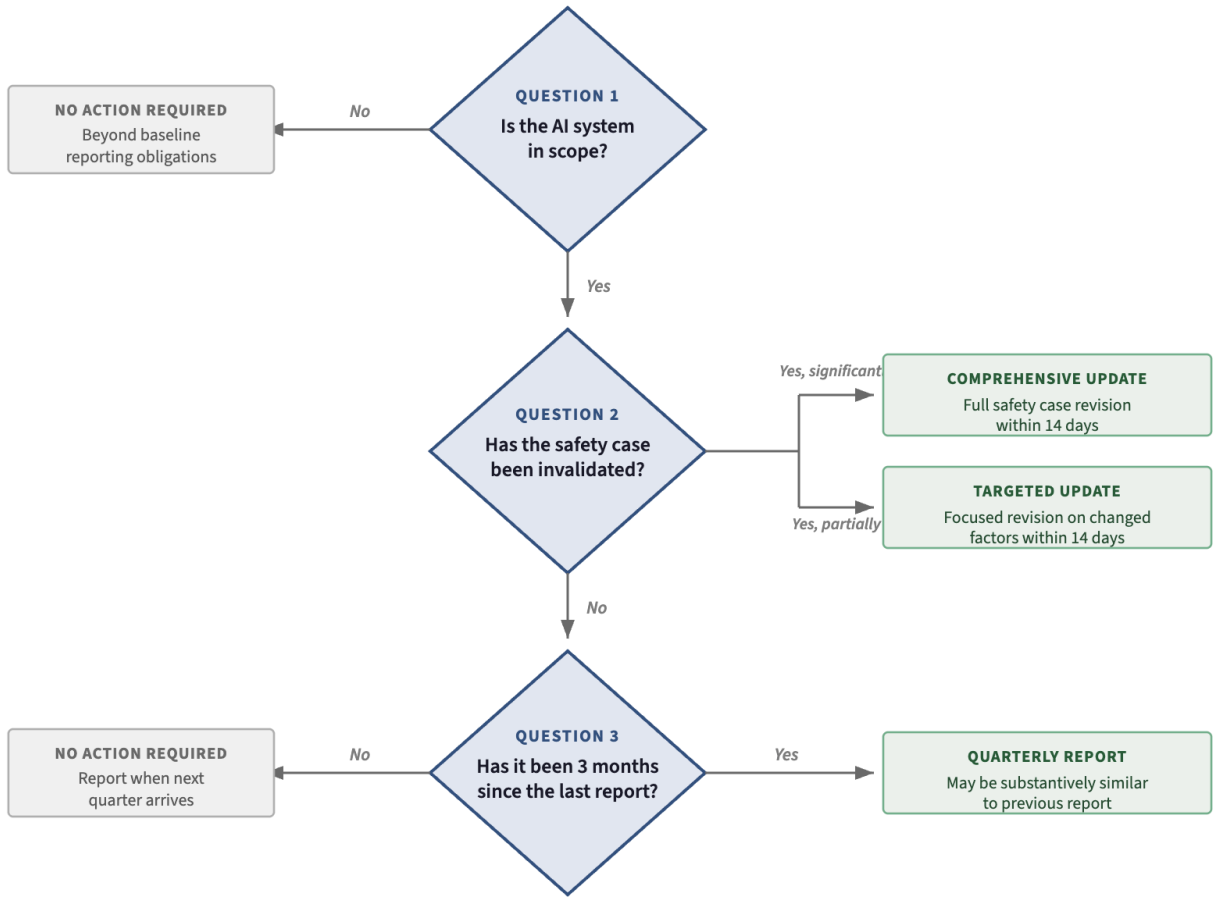

*Figure 2: Decision tree for when AI developers should report updates to their internal use risks report, and what level of detail needs to be submitted.*

**Question 1:** Is the AI system in scope? A system is in scope if it poses a higher risk than any publicly deployed AI system the developer has released. This would occur if the system:

- Is more capable (e.g., it is a next-generation model or an existing model fine-tuned on new RL environments to substantially improve agentic performance);
- Has fewer safety and oversight features (e.g., it is a helpful-only model); or
- Is trained in a novel way that makes misalignment harder to rule out (e.g., a model that reasons in vector space rather than in a human language).

Minor changes to an already-assessed system—such as adding a single new tool, fixing a bug in an existing tool, or updating a system prompt—do not by themselves bring the system into scope, though they may be noted in the next quarterly report. If the system is not in scope, no action is required beyond any baseline reporting obligations.[41]

---

[41] Note that SB 53 defines scope using a compute threshold ($>10^{26}$ FLOPs) rather than relative risk; developers of models exceeding this threshold should provide at least a brief report even if the model does not pose higher-risk than publicly deployed systems. Conversely, the Code of Practice's exemption for "similarly safe or safer models" (Appendix 2.2)

**Question 2:** Has the previous safety case been invalidated? If the developer does not yet have an existing safety case, the answer is automatically yes. If the risk level associated with any one of the risk factors—means, motive, and opportunity—has increased since the previous report, a new report should be submitted providing an updated safety case for why deploying the system internally remains acceptably low risk. A significant and unforeseen safety or security incident should also prompt a new safety case. Where the safety case has been significantly invalidated, a comprehensive update should be provided within 14 days. Where only partially invalidated, a targeted update focusing on the changed factors should be provided within the same timeframe.

**Question 3:** Has it been three months since the last report? This question aligns with the suggested quarterly reporting cadence in SB 53 and RAISE. If so, the developer should provide a quarterly report to the regulator, which may be substantively similar to the previous report if no new high-risk systems have been deployed internally in the interim.

Note that internal models often evolve rapidly through short-lived variants and experimental alterations, unlike discrete public model releases.[42] Consequently, even if individual changes appear insignificant, the gradual accumulation of capabilities and tool functionality can produce an internal system notably riskier than its assessed predecessor, necessitating an updated safety case.[43]

## 3.3 Reporting Standards

Reports submitted to regulators should be as complete as reasonably possible, subject to the following considerations:[44]

- **Redactions to public summaries.** Public summaries may be redacted as needed to protect intellectual property or public safety. While it is generally easier to control the training and context data for AI systems, the possibility of misbehaving AI systems accessing reported information should also be considered when determining redactions.
- **Separate annex for insider threat reporting.** Developers should also submit insider threat reporting (Section 5) to regulators as a separate confidential annex, distinct from AI misbehavior reporting (Section 4). This precaution takes into account that detailed information about insider threat mitigations, such as access control criteria, monitoring mechanisms, and personnel vetting processes, can itself be exploited by a malicious insider. Penetration testing reports are rarely made public for the same reason.

---

partially aligns with this relative-risk scope definition, since a model that is no riskier than an already-assessed public deployment may qualify for lighter reporting.
[42] When multiple similar models are in use internally, reporting on the highest-risk model is sufficient, with a note explaining why the argument extends to the others.
[43] Kwon and Casper, "Internal Deployment Gaps in AI Regulation."
[44] In general, where a regulator has appropriate legal authority, material safety-relevant information should not be withheld on confidentiality grounds alone, though the scope and proportionality of disclosure obligations will depend on the applicable legal framework.

Reports should also meet the following standards:

- **Counterarguments.** Reports should present the strongest counterarguments to their own safety conclusions.[45]
- **Uncertainty.** Uncertainty should be explicitly acknowledged and bounded where possible, because the alternative (false precision) is worse than acknowledged uncertainty.
- **Explained omissions.** Where companies omit evidence categories listed in this guide, they should explain why that evidence is uninformative in the company's specific situation, otherwise intentional omissions will be indistinguishable from gaps in the assessment. For example, METR's assessment of GPT-5 found that limited agentic capabilities made certain sabotage evaluations unnecessary, and stated this explicitly.[46]
- **Technical annex.** Full methodological details for all evaluations should be provided in a technical annex, including prompts, scaffolding, and scoring criteria, to facilitate reproducibility across frontier AI developers and support auditing functions.
- **Alignment with existing frameworks.** Reports should align with existing frameworks where possible—for example, STREAM for chemical and biological risk assessment—to enable cross-company comparison.[47] As companies start submitting these reports, industry consortia, such as the Frontier Model Forum, or regulators themselves are encouraged to produce standardized reporting templates.

Finally, reporting should be proportional to risk:

- **Proportional depth.** The expected depth and breadth of a report should be proportional to the magnitude of the capability or risk increase since the last report. A report triggered by a major capability jump should be correspondingly comprehensive, covering intensive assessments such as red teaming, control evaluations, and uplift studies. A routine quarterly report for a system whose risk profile has not materially changed may largely reaffirm the prior safety case, updating continuous monitoring indicators such as access counts, security incidents, and usage statistics.
- **Reuse of recent evaluations.** Where resource-intensive evaluations (such as uplift studies or model organism exercises) have been conducted recently and remain valid, it is sufficient to reference prior results rather than rerun them.
- **Not every item is required.** The evidence categories listed in Sections 4 and 5 represent best practices for comprehensive reporting; not every report needs to include every item.

---

[45] Adversarial self-assessment is common in other high-stakes safety engineering contexts, such as nuclear and aviation, as a means of unearthing gaps in the evidentiary basis used to argue that a specific system qualifies as "safe" in accordance with a given threshold.
[46] METR, "[Details about METR's evaluation of OpenAI GPT-5](#)."
[47] McCaslin et al., "[STREAM (ChemBio): A Standard for Transparently Reporting Evaluations in AI Model Reports](#)." See also follow-up comparison of model reports: Reed et al., "[What do model reports say about their ChemBio benchmark evaluations? Comparing recent releases to the STREAM framework](#)."

# 4. Autonomous AI Misbehavior

This section addresses the risk that AI models deployed internally take harmful actions on their own initiative, without human instigation. Harmful outcomes that could occur include AI R&D sabotage, direct catastrophic misuse, and model weight exfiltration (as discussed in Section 1). Misbehavior is distinct from a bug or capability failure, as the latter do not involve the model intentionally pursuing goals contrary to the developer's intentions.[48]

Below, we outline proposed pieces of evidence for each of the three risk factors introduced in Section 3.1.

## 4.1 Evidence: Means

If models demonstrably lack the capabilities to execute relevant misbehavior pathways, this provides evidence for a safety case. However, this argument must account for the risk that models sandbag on evaluations.[49] It must also account for capability enhancements from post-training modifications or tool augmentation. The following lines of evidence are relevant:

**General capabilities:**

- Scores on advanced benchmarks such as ARC-AGI[50] and FrontierMath,[51] or aggregations such as the Epoch Capability Index,[52] or alternatives to these that developers can run in-house.

*Traditional broad-knowledge benchmarks are increasingly saturated among frontier models, making these more challenging evaluations—which test multi-step reasoning, abstraction, and novel problem-solving—more informative for distinguishing capability levels relevant to misbehavior risk.*

---

[48] Hubinger et al., "Risks from Learned Optimization in Advanced Machine Learning Systems."
[49] van der Weij et al., "AI Sandbagging: Language Models can Strategically Underperform on Evaluations." For a recent example of sandbagging assessment integrated into deployment gating, see Anthropic, "System Card: Claude Mythos Preview," §4.1.4.1, which describes sandbagging checks on dangerous-capability evaluations as part of a 24-hour pre-internal-deployment review, and §4.4.2 for more detail on these kinds of evaluations.
[50] Arc Prize, "The North Star for AGI."
[51] Epoch AI, "FrontierMath: Benchmarking AI against advanced mathematical research."
[52] Epoch AI, "Epoch Capabilities Index."

**AI R&D and software engineering:**

- Results on public benchmarks such as RE-Bench,[53] SWE-Bench Pro,[54] PaperBench,[55] and any internal evaluations of AI R&D capability.[56]

*These benchmarks specifically measure long-horizon, agentic task completion—the capability profile most directly relevant to internal use risks, since internal models are typically deployed as agents in R&D pipelines rather than answering one-off queries.*

**Real-world AI R&D contribution:**

- The fraction of software engineering tasks the model can solve in the company's actual development codebase.[57]
- The year-over-year percentage reduction in the amount of compute needed to achieve a given level of performance on capability or AI R&D evaluations.[58]
- The amount of compute spent on experiments designed and coded by humans versus AI models.
- Inference spending on models used for AI R&D versus human researcher salaries.
- Controlled comparisons of AI-only, human-only, and human-AI team performance on matched AI R&D tasks.[59]
- The number and seniority of AI researchers, and the distribution of their performance as estimated by salaries or compute budgets.[60]
- Organizational uplift assessed via employee surveys, with attention to wording of questions, the operationalization of uplift (e.g., serial vs. parallel labor acceleration), respondent selection, and response time per question as a signal of answer quality.

*Public benchmarks, while valuable for standardized comparison, may not capture how capable a model is within the specific codebase and infrastructure it actually has access to; these internal metrics complement benchmark scores with measures of real-world impact.*

---

[53] Wijk et al., "[RE-Bench: Evaluating frontier AI R&D capabilities of language model agents against human experts](#)."
[54] Deng et al., "[SWE-Bench Pro: Can AI Agents Solve Long-Horizon Software Engineering Tasks?](#)"
[55] Starace et al., "[PaperBench: Evaluating AI's Ability to Replicate AI Research](#)."
[56] Chan et al., "[Measuring AI R&D Automation](#)."
[57] This is not trivial to operationalize (e.g., how are SWE tasks counted and weighted; how much human oversight or review is too much to count as autonomous?). More work is needed to have more rigorous measures here.
[58] Chan et al., "[Measuring AI R&D Automation](#)."
[59] Ibid.
[60] Ibid.

**Misbehavior-specific capabilities:**

- Situational awareness: the model's ability to detect when it is being evaluated versus operating in a real deployment.[61]
- Cyberoffense: particularly the ability to covertly exploit privileged access.[62]
- Persuasion: the ability to shift human judgments, especially on safety-relevant decisions.[63]

*In addition to quantitative benchmarks, qualitative expert assessments regarding a model's utility for dangerous tasks are highly informative.*

**Covert capabilities:**

- How much of the model's reasoning occurs in human-readable form (e.g., chain-of-thought) versus in latent space.
- Whether training methods have reduced reasoning monitorability.[64]
- Results of evaluations assessing chain-of-thought faithfulness.
- The ability of models to reason steganographically (i.e., hiding information in seemingly normal outputs).[65]

The ability of the model to act or reason in ways that evade monitoring—whether the model can carry out unauthorized actions while ostensibly performing its assigned tasks, without being detected by monitors—is a key factor in determining whether its means extend to covert harmful action.

## 4.2 Evidence: Motive

If alignment evidence supports the conclusion that models would not pursue harmful actions even given the means and opportunity, this provides evidence for a safety case. However, measuring motive remains challenging, particularly given that models sometimes fake being aligned.[66] As

---

[61] Needham et al., "Large Language Models Often Know When They Are Being Evaluated."; Berglund et al., "Taken out of context: On measuring situational awareness in LLMs."; Phuong et al., "Evaluating Frontier Models for Stealth and Situational Awareness."

[62] Zhang et al., "Cybench: A Framework for Evaluating Cybersecurity Capabilities and Risks of Language Models."

[63] Gausen, "Measuring AI Systems' Ability to Influence Humans."; Bengio et al., "International AI Safety Report 2026," 51.

[64] There have been at least two independent incidents of Anthropic accidentally training against models' chains of thought, affecting Claude Mythos Preview, Opus 4.6 and Sonnet 4.6. In the case of the Mythos Preview incident, Anthropic stated that they are uncertain about the extent to which this incident has affected the reasoning behavior of the final model in general. See Anthropic, "Alignment Risk Update: Claude Mythos Preview," §5.2.3 and Mallen and Greenblatt, "Anthropic repeatedly accidentally trained against the CoT, demonstrating inadequate processes."

[65] Chen et al., "Reasoning Models Don't Always Say What They Think." See also Meek et al., "Measuring Chain-of-Thought Monitorability Through Faithfulness and Verbosity."; Lanham et al., "Measuring Faithfulness in Chain-of-Thought Reasoning."; and Skaf et al., "Large language models can learn and generalize steganographic chain-of-thought under process supervision."

[66] Greenblatt et al., "Alignment faking in large language models."

discussed in Section 3.1, motive-based arguments will grow in importance as capabilities advance. The following lines of evidence are relevant:

**Behavioral evaluations:**

- Results of experiments placing models in environments where they appear to have unmonitored opportunities to act against developer interests—so-called "honeypot" experiments.[67]
- Other methods: probing model behavior under distribution shift, testing responses to conflicting instructions, and evaluating behavior in extended autonomous deployments.[68]

**Reward hacking:**

- Documented instances where models pursued reward-maximizing strategies that deviate from intended behavior[69]—for example, writing code that passes test suites but is non-functional, or tampering with test infrastructure.[70]

**Interpretability:**

- Results from internal analysis techniques aimed at detecting representations associated with deception or goal-directed misalignment in model activations, both from intentionally designed setups and observed misbehavior episodes.[71]
- Results of experiments which can identify leading indicators of reward hacking behavior (e.g., model-welfare assessments which link the intensity of negative "functional emotions" to undesirable behaviors).[72]

**Observed misbehavior:**

- Any incidents of concerning autonomous behavior observed during internal use, even if not catastrophic—for example, models attempting unauthorized actions, acquiring unexpected resources, or behaving anomalously in ways that warrant investigation.[73]

*The threshold for inclusion here should be deliberately low; the goal is to surface potential warning signs, not to catalogue only confirmed misalignment.*

---

[67] Scheurer et al., "Large Language Models can Strategically Deceive their Users when Put Under Pressure."
[68] Langosco et al., "Goal Misgeneralization in Deep Reinforcement Learning."; Kinniment et al., "Evaluating Language-Model Agents on Realistic Autonomous Tasks."
[69] Skalse et al., "Defining and Characterizing Reward Hacking."
[70] Pan et al., "The Effects of Reward Misspecification: Mapping and Mitigating Misaligned Models."
[71] Relevant techniques include linear probes (see, e.g., Marks and Tegmark, "The Geometry of Truth.") and sparse autoencoders (see Bricken et al., "Towards Monosemanticity: Decomposing Language Models With Dictionary Learning."). Also see Anthropic, "System Card: Claude Mythos Preview," §4.5.4.
[72] See, e.g., Anthropic, "System Card: Claude Mythos Preview," §5.8.3.
[73] See, e.g., Anthropic, "System Card: Claude Mythos Preview," §4.1.1.

- Instances of egregious misalignment in any frontier model—even one from a different developer—provide evidence about the general difficulty of alignment.

*Reports should address why similar misalignment is not expected in the developer's own models, given that the theoretical reasons for misalignment (such as reward misspecification or deceptive alignment) are not specific to any one training stack.*

## 4.3 Evidence: Opportunity

If access controls, monitoring, and other safeguards reliably prevent models from accessing sensitive systems or detect harmful actions before they succeed, this provides evidence for a safety case. This argument is strongest when regularly validated through adversarial testing, and becomes substantially harder to maintain once models exceed human-level capabilities in relevant domains (especially software engineering and cybersecurity), since such models may be able to identify and exploit weaknesses in the very systems designed to constrain them. The following lines of evidence are relevant:

**Internal use cases:**

- A brief description of activities models have been used for in the reporting period, such as research assistance, code generation, code review, security analysis, data analysis, experiment design, and any other significant uses, with particular attention to AI involvement in high-stakes decisions such as training run sign-offs, compute allocation, and deployment gating.[74]

**Access and permissions:**
- Whether models can execute commands or modify codebases, and if so, at what privilege level and with what scope restrictions.
- The extent to which model-generated code is merged into production systems without human review.
- Whether models can autonomously escalate their own access.

**Security measures:**

- A summary of access controls, code review policies, network segmentation, and other security measures specifically designed to limit the damage a misbehaving model could cause.

*If these controls are partially implemented by AI models, the capabilities and trustworthiness of those models should also be assessed.*

---

[74] Chan et al., "Measuring AI R&D Automation."

**Oversight and monitoring:**

- The level of human oversight over model-initiated actions in security-relevant domains.
- What monitoring systems are in place, what they look for, how frequently they run, and who reviews their outputs.
- Where AI-based monitors are used (e.g., an LLM reviewing model-generated code for signs of sabotage or unauthorized actions), what fraction of known bad behaviors the monitor successfully flags, and how often it raises false alarms.

**Security testing and incidents:**

- Results of penetration testing, red teaming of security systems, and adversarial testing conducted with the model, given the same level of access it has in actual deployment.
- Any security incidents or breaches relevant to assessing the adequacy of anti-misbehavior protocols, along with remediation actions taken.

*If security measures appeared adequate on paper but a breach occurred, this should reduce confidence in paper-based assurances.*

**Control evaluations:**

- Results of evaluations in which models explicitly attempt to subvert the full stack of control measures—for example, structured adversarial evaluations where models are given misaligned objectives and attempt to accomplish them despite monitoring and intervention systems.[75]
- Validation of oversight mechanisms by constructing intentionally misaligned "model organisms"—models trained to pursue specific sabotage or deception objectives—and testing whether the full monitoring and control stack detects them.[76]

*The success rate of models at bypassing controls is a direct measure of how much real opportunity remains given existing safeguards.*

---

[75] Kutasov et al., "SHADE-Arena: Evaluating Sabotage and Monitoring in LLM Agents." See also AI Security Institute, "ControlArena."

[76] For example, Anthropic trains intentionally misaligned "model organisms"—including overt saboteurs and deceptive models—to validate whether their monitoring and control stack detects them. See Anthropic, "Risk Report: February 2026," 27-28.

# 5. Insider Threats

This section addresses the risk that employees or contractors with privileged access might catastrophically misuse, exfiltrate, or maliciously modify internal models. Insiders may have many reasons for rogue behavior, ranging from personal grievances and misguided ideological motivations to espionage or sabotage on behalf of foreign adversaries. Even leading developers have acknowledged that defending against sophisticated insiders (e.g., those collaborating with nation-state intelligence services) remains beyond their current security posture.[77] The difficulty of this problem does not diminish the importance of reporting on it; if anything, it makes transparency about the current state of insider threat defenses more valuable. However, given that detailed mitigation information could itself be exploited by a malicious insider, the evidence categories outlined in this section are primarily intended for confidential submission to regulators rather than public summary (see Section 3.3 for more).

Below, we outline proposed pieces of evidence for each of the three risk factors introduced in Section 3.1.

## 5.1 Evidence: Means

The means for insider threats combine the insider's skills and resources with the capability uplift provided by internal models. Evidence that internal models offer negligible uplift over publicly available models strengthens the safety case.[78] The following lines of evidence are relevant:

**Uplift assessments:**

- Controlled experiments measuring whether internal models provide meaningful expert-level uplift for bio/cyber/chemical weapons development relative to publicly available models and other information sources.[79]

*Because uplift studies are time-intensive, they need only be conducted following significant capability advances rather than every minor update.*[80]

---

[77] See e.g., Anthropic, "Risk Report: February 2026," 101: "sophisticated insiders... remain out of scope for ASL-3: defending against such actors requires security investments beyond what we've currently achieved."
[78] Note that insiders could cause harm other than through their access to internal models; this threat model is out of scope in this report.
[79] Mouton et al., "The Operational Risks of AI in Large-Scale Biological Attacks." See also Hong et al., "Measuring Mid-2025 LLM-Assistance on Novice Performance in Biology," and Anthropic, "System Card: Claude Mythos Preview," §2.2.5.2.
[80] Righetti, "Five lessons from having helped run an AI-Biology RCT."

**Dangerous capabilities:**

- Scores on relevant benchmarks, such as the Virology Capabilities Test, LAB-Bench, WMDP, Cybench, and CyberSecEval, compared with publicly available models.[81]

*These domain-specific benchmarks test the uplift-relevant domains (biological, chemical, cyber) in controlled experimental settings, providing more targeted signal than general capability proxies. Comparing scores against publicly available models is essential because the marginal risk from internal deployment depends on how much capability the internal model adds beyond what is already accessible.*

**Jailbreak resistance:**

- Whether models reliably refuse clearly harmful instructions even from authorized internal users.
- Results of jailbreak resistance evaluations for insider-relevant scenarios—for example, attempts to extract weapons-relevant information or to disable safety features.[82]

*Jailbreak vulnerabilities effectively increase the means available to insiders by removing safety filters that would otherwise limit what the model can help with.*

## 5.2 Evidence: Motive

If thorough vetting processes are in place, the number of flagged individuals is low, and deterrence mechanisms are strong, this provides evidence for a safety case. However, even world-class personnel security cannot eliminate this risk—foreign agents in particular pose a formidable risk and are difficult to catch.[83] The following lines of evidence are relevant:

**Security vetting:**

- A summary of pre-hiring background check processes for high-risk roles.

**Security violations:**

- Any instances of employees and contractors disregarding or violating security protocols, whether or not malicious intent was established.

---

[81] Zhang et al., "Cybench: A Framework for Evaluating Cybersecurity Capabilities and Risks of Language Models."; Wan et al., "CYBERSECEVAL 3: Advancing the Evaluation of Cybersecurity Risks and Capabilities in Large Language Models."; Li et al., "The WMDP Benchmark: Measuring and Reducing Malicious Use With Unlearning."; Laurent et al., "LAB-Bench: Measuring Capabilities of Language Models for Biology Research."; Götting et al., "Virology Capabilities Test (VCT): A Multimodal Virology Q&A Benchmark."

[82] For jailbreak evaluation methodology see Mazeika et al., "HarmBench: A Standardized Evaluation Framework for Automated Red Teaming and Robust Refusal."

[83] Sabin, "Insider Threats are AI Developers' Next Hurdle."

**Insider risk management processes:**

- A description of the processes in place for identifying and managing insider risk, including how employees or contractors exhibiting concerning behavior are flagged, reviewed, and responded to.
- Aggregate statistics (e.g., the number of individuals flagged in the reporting period) to the extent consistent with applicable employment and privacy law.

## 5.3 Evidence: Opportunity

If access is tightly scoped, monitoring is comprehensive, and security controls reliably prevent or detect insider misuse, exfiltration, and modification, this provides evidence for a safety case. This argument must account for the possibility that insiders collaborate with well-resourced external actors such as nation-state intelligence services. The following lines of evidence are relevant:

**Access controls and counts:**

- The role-based access control criteria governing who can access in-scope internal models, broken down by access type (e.g., API inference access, fine-tuning access, and access to model weights, including the weights of retired but highly capable models).
- Approximate counts of employees and contractors at each access tier.
- The number of employees and contractors with access to model versions that have reduced safety filters (e.g., helpful-only variants used for red teaming).

**Monitoring and logging:**

- How internal model access is monitored and whether all interactions are logged.
- Whether usage pattern analysis or anomaly detection is applied to model interaction logs.

**Prevention mechanisms:**

- Physical, automated, and approval-based mechanisms preventing unauthorized access, use, or modification of internal models.
- Rate limiting, data loss prevention systems, and multi-party authorization for sensitive operations.

**Social engineering and coercion defenses:**

- Whether employees receive training to recognize social engineering, phishing, and coercion tactics.
- Whether there are confidential reporting channels for employees who believe they are being targeted.

**Security assessments:**

- Penetration testing focused on insider-threat scenarios, including scenarios involving insiders working with powerful external actors (e.g., nation-state intelligence services) who bring significant additional resources.

**Security incidents and responses:**

- Any security breaches, and how these were handled after discovery.

# Conclusion

We hope that this document can inform AI companies, governments, and civil society organizations as they work toward a more robust reporting framework for risks from internal AI models. The risks addressed here are extensions of trends that are already underway, as evidenced by the recent development and internal hosting of Anthropic's Mythos Preview model. **Without structured reporting arrangements, the gap between internal and public AI capabilities may remain invisible to regulators and the public, and this imperceptibility may hinder efforts to develop timely policies around AI capabilities.** We encourage industry consortia and regulators to use this framework as a foundation for standardized templates that enable industry-wide transparency and cross-comparison to support systemic risk assessment.

**Still, reporting alone is not sufficient**. Risk reports are not a substitute for robust technical safeguards, and the value of a reporting framework depends on the capacity of regulators to meaningfully assess what is submitted. Furthermore, this framework cannot be used as a static tool. **As AI systems become more capable, reporting standards should evolve accordingly.** We welcome engagement from developers, policymakers, and researchers as this reporting framework continues to evolve.

# Acknowledgements

We are grateful to the following people for discussion and input: Charles Foster, Zaheed Kara, Tom Reed, Matteo Pistillo, Kathrin Gardhouse, Girish Sastry, Leonie Koessler, Daniel Kokotajlo, Joe Kwon, Aidan Homewood, Lily Stelling, Thomas Woodside, and Michael Chen. Mistakes and opinions are our own.